\documentclass[conference]{IEEEtran}
\IEEEoverridecommandlockouts
\usepackage{cite}
\usepackage[caption=false,font=footnotesize]{subfig}

\usepackage{amsmath,amssymb,amsfonts}
\usepackage{booktabs} 
\usepackage{algorithmic}
\usepackage{graphicx}
\usepackage{textcomp}
\usepackage{xcolor}
\usepackage{enumitem}
\usepackage{multicol}
\usepackage{multirow}
\usepackage{hyperref}
\usepackage{xurl}
\usepackage{soul}
\def\BibTeX{{\rm B\kern-.05em{\sc i\kern-.025em b}\kern-.08em
    T\kern-.1667em\lower.7ex\hbox{E}\kern-.125emX}}
\begin{document}

\title{%Multidimensional Resilience Assessment of Power Systems under Coordinated Cyber-Physical Attacks\\
GRID-MRI: Resilience Assessment of Power Systems under Coordinated Cyber-Physical Attacks \\

%\thanks{This work is funded by the National Science Foundation (NSF) Award Number \#2501975.}
}

\author{\IEEEauthorblockN{Isaac Ortega Romero and  Ioannis Zografopoulos}
\IEEEauthorblockA{\textit{Engineering Department, University of Massachusetts Boston, Boston, MA, USA} \\
%\textit{University of Massachusetts Boston}\\
% Boston, MA, USA \\
\textit{email: \{I.OrtegaRomero001 and I.Zografopoulos\}@umb.edu} \vspace{-5mm}
%0009-0000-6627-6649
}}

\maketitle

\begin{abstract}
The increasing penetration of inverter-based resources and growing cyber-physical interdependencies expose electrical power systems to disturbances that challenge conventional resilience assessments. Existing approaches often evaluate resilience dimensions independently, overlooking the non-additive effects of coordinated disruptions and the influence of exogenous stressors. To address this gap, this paper proposes a Grid Multidimensional Resilience Index (GRID-MRI) that integrates endogenous disruption severity, coordination effects, and climatic and economic-regulatory conditions. The framework quantifies physical, cyber, and operational degradation, identifies interaction effects through counterfactual (i.e., what-if),  simulations of individual and coordinated attacks, and incorporates exogenous stressors as scenario-dependent scaling factors. Its application to a modified IEEE 39-bus system under three escalating cyber-physical attack scenarios, informed by the December 2025 Polish cyberattack incident, reveals that coordinated attacks induce system collapse despite stable individual and additive responses. The corresponding resilience losses are 3.72 and 4.09 times their additive estimates, while GRID-MRI increases from 0.062 to 1.188 and reaches 6.081 under additional exogenous stress. These findings demonstrate the limitations of additive assessments and establish a quantitative basis for distinguishing the dominant mechanisms of multidimensional resilience degradation.
\end{abstract}


\section{Introduction} \label{s:Intro}

The increasing integration of inverter-based resources and digital control infrastructure is transforming electrical power systems, introducing new interdependencies and vulnerabilities to coordinated cyber-physical disturbances. Under high-impact, low-probability events, resilience depends not only on the condition of individual assets but also on the ability of interconnected systems to withstand degradation and maintain essential functions~\cite{Stankovic2023,KellyGorham2024,Wang2022}. As physical infrastructure, operational decisions, and digital control become increasingly interdependent, disturbances affecting multiple assets may propagate through the system and produce consequences that cannot be inferred from isolated failures.

Existing research has advanced resilience assessment through performance-based metrics and degradation-recovery curves~\cite{Dobson2023,Talukder2021,Mujjuni2021}, multi-stage evaluation frameworks~\cite{Stankovic2023,Wang2025,YooPark2024}, and models incorporating infrastructure interdependencies and compound hazards~\cite{KellyGorham2024,Ti2022,Sun2024}. However, these approaches commonly emphasize specific performance indicators, individual resilience dimensions, or selected coupling mechanisms. Consequently, such metrics may characterize disruption severity without explicitly distinguishing the effects of individual failures from the additional degradation caused by their coordination. This limitation becomes critical when individually tolerable disturbances collectively induce system-wide instability. Beyond endogenous interactions, external conditions can further exacerbate the consequences of cyber-physical disruptions. Extreme weather may simultaneously increase electricity demand, stress infrastructure, and constrain restoration resources, reducing the system's capacity to absorb additional disturbances~\cite{Wang2022,Ti2022, zografopoulos2022mitigation}. Similarly, insufficient preventive investment and regulatory oversights can weaken cybersecurity preparedness and prolong recovery, increasing an attack's economic consequences. 
%The separate analysis of these factors further limits their integration with dynamic system responses within a resilience assessment under compound disturbances. 
\textcolor{black}{Analyzing factors impeding resilience independently further limits their potential to be utilized to inform system responses and can also underestimate the comprehensive impact of disturbances.}

To address these gaps, this paper proposes a \textcolor{black}{Grid Multidimensional Resilience Index (GRID-MRI)} that integrates five resilience dimensions: \emph{i)} physical, \emph{ii)} cyber, and \emph{iii)} operational, representing endogenous disruptions and \emph{iv)} climatic and \emph{v)} economic-regulatory, representing exogenous factors. The index distinguishes aggregate disruption severity, non-additive coordination effects, and exogenous amplification. The main contributions are as follows:

\begin{itemize}%[leftmargin=*]
    \item The \textcolor{black}{GRID-MRI} formulation that separates endogenous disruption severity, non-additive coordination effects identified through counterfactual simulations, and scenario-dependent climatic and economic-regulatory amplification.
    \item The application of GRID-MRI on a modified IEEE 39-bus system under three escalating cyber-physical attack scenarios informed by the December 2025 Polish incident~\cite{CERTPolska2026}, demonstrating system collapse under coordinated attacks despite stable individual and additive responses.
\end{itemize}

The remainder of this paper is organized as follows: Section~II presents the multidimensional resilience assessment framework and \textcolor{black}{GRID-MRI} formulation, Section~III describes the case study and results, and Section~IV concludes the paper.

\section{Multidimensional Resilience Assessment Framework} \label{s:method}

\textcolor{black}{The framework in~\cite{romero2026resilience} accounted for cross-dimensional interactions through a predefined algebraic coupling term, a limitation addressed in the present study by experimentally quantifying non-additive coordination effects through counterfactual, i.e., what-if worst-case scenario, simulations.}

\subsection{System Performance and Resilience Loss}

For each scenario $S_i$, system performance is quantified by the equally weighted contributions of: \emph{i)} centre-of-inertia (COI) frequency deviation and \emph{ii)} inter-generator frequency dispersion, both evaluated relative to the corresponding non-attacked baseline, defined as:

\vspace{-15pt}
\begin{align}
\Phi_i(t)&=\tfrac12\bigl(\Phi_{f,i}(t)+\Phi_{s,i}(t)\bigr),\label{eq:performance}\\
\Phi_{f,i}(t)&=\left[1-\frac{|\Delta f_{\mathrm{COI},i}(t)|}{1\,\mathrm{Hz}}\right]_+,\\
\Phi_{s,i}(t)&=\left[1-\frac{\Delta f_{\mathrm{gen},i}(t)}{1\,\mathrm{Hz}}\right]_+,
\end{align}
where $[x]_+=\max(0,x)$, $\Delta f_{\mathrm{COI},i}(t)=f_{\mathrm{COI},i}(t)-f^0_{\mathrm{COI}}(t)$, and $\Delta f_{\mathrm{gen},i}(t)=\max_{j,k}|\Delta f_{j,i}(t)-\Delta f_{k,i}(t)|$, with $\Delta f_{j,i}(t)=f_{j,i}(t)-f_j^0(t)$. The two components capture system-wide frequency deviations and generator coherency degradation, respectively.

To distinguish performance degradation from system collapse, the collapse time $t_{\mathrm{col},i}$ is defined as the first post-attack instant at which $\Delta f_{\mathrm{gen},i}(t)>1$~Hz or $\Delta f_{\mathrm{COI},i}(t)<-1$~Hz, thresholds distinct from the performance normalization limits. Accordingly, $\widetilde{\Phi}_i(t)=\Phi_i(t)$ before collapse and $\widetilde{\Phi}_i(t)=0$ for $t\geq t_{\mathrm{col},i}$; in the absence of collapse, $\widetilde{\Phi}_i(t)=\Phi_i(t)$ throughout the evaluation horizon. The resulting resilience loss is:

\vspace{-3pt}
\begin{equation}
R_{\mathrm{loss},i}=\frac{1}{\Phi_{0,i}}\int_{T_0}^{T_0+T_H}
\!\bigl[\Phi_{0,i}-\tilde{\Phi}_i(t)\bigr]_+\,dt,
\label{eq:rloss}
\end{equation}
where $T_0$ denotes the earliest attack instant across scenarios, $T_H$ is the evaluation horizon, and $\Phi_{0,i}$ is the mean performance over the preceding second. Normalization by $\Phi_{0,i}$ expresses the cumulative degradation relative to pre-attack performance, yielding an equivalent duration of complete performance loss.

\subsection{Multidimensional Disruption Indices}

The disruption indices characterize: \emph{i)} physical, \emph{ii)} cyber, \emph{iii)} operational, \emph{iv)} climatic, and \emph{v)} economic-regulatory impacts complementary to the resilience-loss metric.

\textit{Physical and cyber:} The respective indices quantify lost generation capacity and compromised digital assets:

\vspace{-10pt}
\begin{align}
D_{\mathrm{phy},i}&=\sum_{r\in\mathcal R}\nu_r\frac{P_{r,\mathrm{lost},i}}{P_{r,\mathrm{total}}},\label{eq:phy}\\
D_{\mathrm{cyb},i}&=\sum_{j\in\mathcal K}w_{\mathrm{cy},j}\frac{N_{\mathrm{compr},j,i}}{N_{\mathrm{scope},j}}.\label{eq:cyb}
\end{align}

The set $\mathcal R$ represents exposed resource types, where $P_{r,\mathrm{lost},i}$ and $P_{r,\mathrm{total}}$ denote lost and installed capacities, respectively. Criticality weights satisfy $\sum_r\nu_r=1$, with $\nu_r=1$ for a single resource type. Similarly, $\mathcal K$ includes five equally weighted cyber aspects: \emph{i)} observability, \emph{ii)} controllability, \emph{iii)} integrity, \emph{iv)} availability, and \emph{v)} recoverability \cite{zografopoulos2020harness, 10202890}. \textcolor{black}{The corresponding $N_{\mathrm{compr},j,i}$ and
$N_{\mathrm{scope},j}$ represent the numbers
of compromised and assessed assets, respectively,
while $w_{\mathrm{cy},j}=0.2$ assigns equal weight
to each cyber aspect.} Based on the scenario-specific threat model, the cyber index captures disruptions to monitoring, control, and recovery functions, even when physical generation capacity remains unaffected.

\textit{Operational:} Frequency-response severity is assessed independently of $\Phi_i(t)$ through two indicators: \emph{i)} the maximum absolute rate of change of frequency (RoCoF) and \emph{ii)} the minimum COI frequency deviation relative to the baseline, defined as:

\vspace{-5pt}
\begin{equation}
D_{\mathrm{op},i}=\tfrac12\left(
\frac{|\mathrm{RoCoF}|_{\max,i}}{1\,\mathrm{Hz/s}}+
\frac{|\Delta f_{\mathrm{nadir},i}|}{1\,\mathrm{Hz}}\right),
\label{eq:op}
\end{equation}
where $\Delta f_{\mathrm{nadir},i}=\min_t\Delta f_{\mathrm{COI},i}(t)$. Both indicators are evaluated from the first attack in scenario $S_i$, at $t_{0,i}$, until $\min(t_{\mathrm{col},i},T_0+T_H)$ \cite{entsoe2023inertia}.

\textit{Climatic:} Winter-driven demand is reflected in the dynamic simulation as per \cite{CERTPolska2026}, while scenario-specific infrastructure stress from ice, snow, wind, and temperature is quantified as:

\vspace{-5pt}
\begin{equation}
D_{\mathrm{clim},i}=\sum_{k\in\mathcal C}w_{c_k}I_{c_k,i},\quad
I_{c_k,i}=\frac{x_{k,i}-x_k^0}{x_k^{\mathrm{ref}}-x_k^0},
\label{eq:clim}
\end{equation}
where $\mathcal C$ denotes the stressor set, with weights satisfying $\sum_kw_{c_k}=1$. The terms $x_{k,i}$, $x_k^0$, and $x_k^{\mathrm{ref}}$ represent scenario, baseline, and design-reference stress levels, respectively. Values $I^c_{k,i}>1$ indicate stress beyond the design reference~\cite{Eurelectric2024,EN1991_Annex_C,Chmielewski2023}. \looseness=-1

\textit{Economic-regulatory:} Economic underinvestment is quantified through a bounded function relating realized failure costs to the minimum investment required for prevention or mitigation, expressed as:

\vspace{-15pt}
\begin{gather}
D_{\mathrm{econ},i}^{\mathrm{und}}=1-\exp\!\left(-\frac{C_i^{\mathrm{tot}}}{C_i^{\mathrm{min}}}\right),\label{eq:econ}\\
C_i^{\mathrm{tot}}=\mathrm{VoLL}\,\mathrm{ENS}_i+C_i^{\mathrm{rep}}
+C_i^{\mathrm{pen}}+C_i^{\mathrm{isl}}\nonumber
\end{gather}
The term $C_i^{\mathrm{min}}$ denotes the minimum preventive or mitigation investment for scenario $S_i$, while $\mathrm{VoLL}$ represents the value of lost load and $\mathrm{ENS}_i$ the unserved energy accumulated throughout restoration. The remaining cost terms account for repairs, regulatory penalties, and proactive or reactive islanding \cite{zografopoulos2025event}.

Regulatory vulnerability reflects the extent to which reference controls present exploitable compliance or implementation gaps in $S_i$. These controls, drawn from IEC and NERC CIP guidelines~\cite{IEC62443,NERCCIP}, span technical safeguards, i.e., access control, patch management, default credentials, network segmentation, operational technology (OT) monitoring, and distributed energy resource (DER) cybersecurity compliance, and organizational and physical practices, i.e., incident response, supply chain risk, physical security, and training and awareness. The economic and regulatory dimensions are integrated through a weighted formulation % that includes their interaction, 
defined as:

\begin{figure*}[!t]
  \centering
  \scalebox{0.92}{%
    \includegraphics[width=0.345\textwidth]{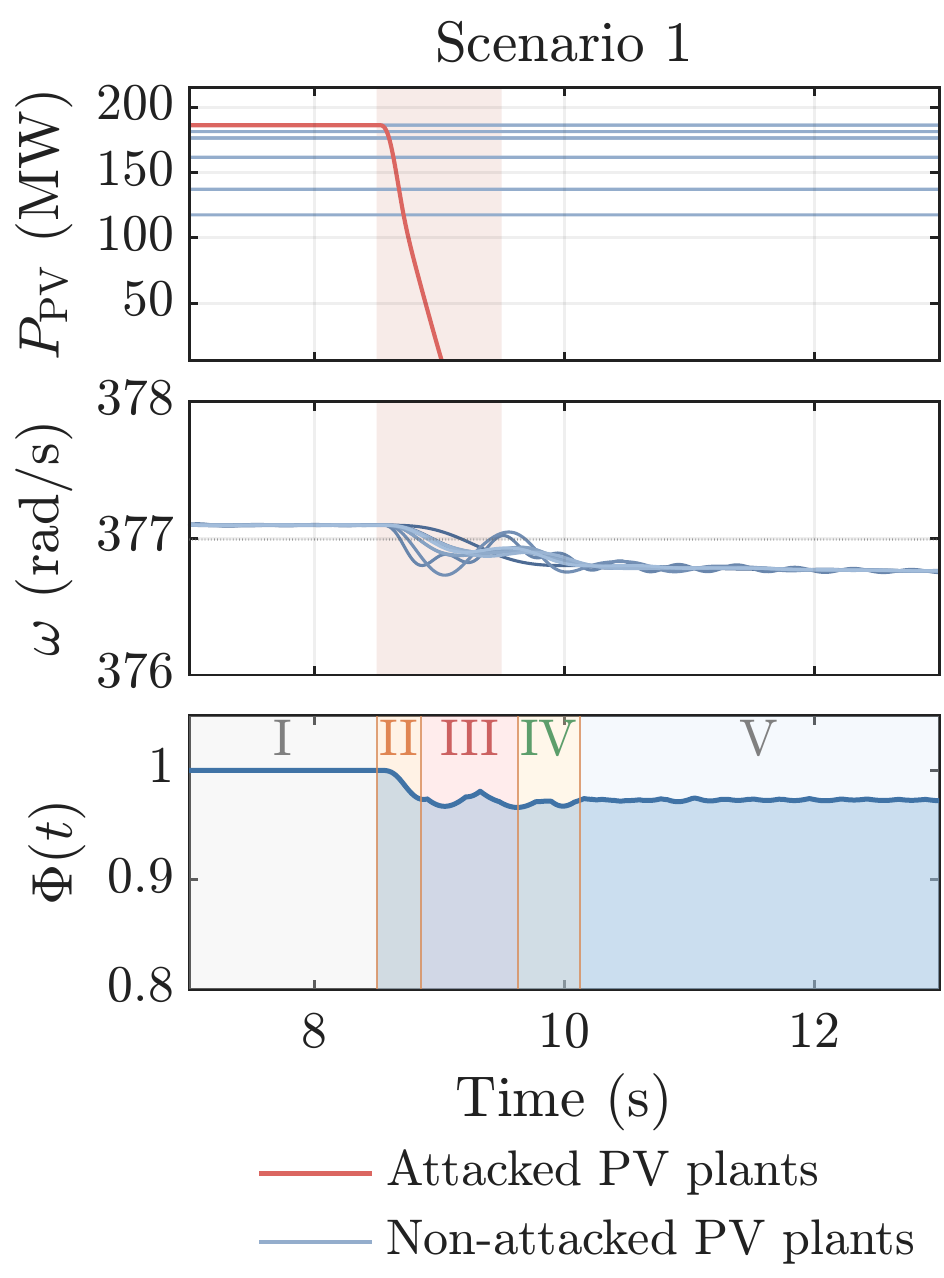}\hspace{0.009\textwidth}%
    \includegraphics[width=0.315\textwidth]{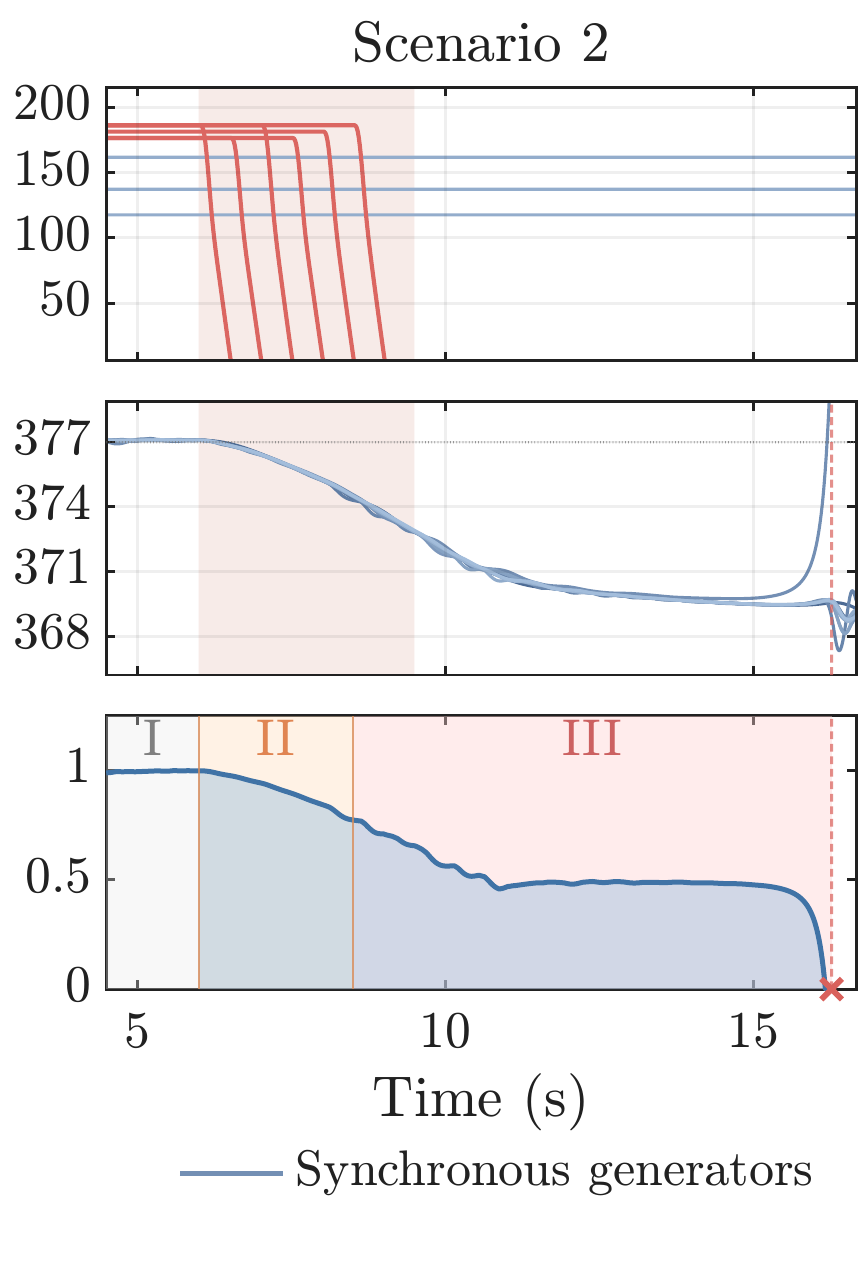}\hspace{0.009\textwidth}%
    \includegraphics[width=0.320\textwidth]{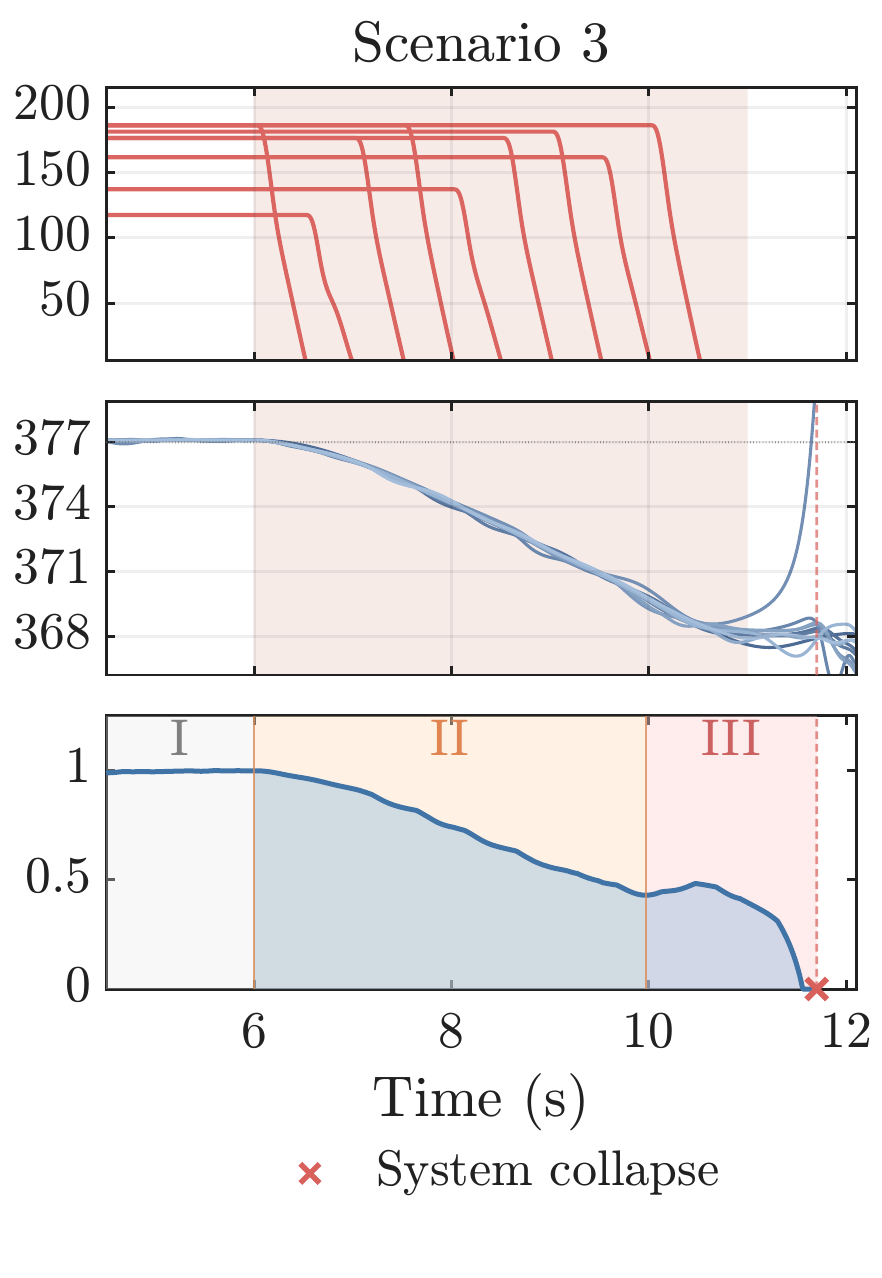}%
  }
\vspace{-3mm}
\caption{Dynamic response of PV generation, synchronous generators, and system performance ($\Phi(t)$) under cyberattacks.}

\label{fig:scenarios} 
\vspace{-4mm}
\end{figure*}

\vspace{-12pt}
\begin{gather}
D_{\mathrm{reg},i}^{\mathrm{vul}}=\frac{1}{N_v^{\mathrm{ref}}}\sum_{j=1}^{N_v^{\mathrm{ref}}}v_{j,i},\label{eq:reg}\\
\resizebox{0.86\columnwidth}{!}{$\displaystyle
D_{\mathrm{econ\text{-}reg},i}=w_eD_{\mathrm{econ},i}^{\mathrm{und}}
+w_rD_{\mathrm{reg},i}^{\mathrm{vul}}
+w_cD_{\mathrm{econ},i}^{\mathrm{und}}D_{\mathrm{reg},i}^{\mathrm{vul}},$}\label{eq:econreg}
\end{gather}
where $N_v^{\mathrm{ref}}=10$ is the number of reference controls and $v_{j,i}\in\{0,1\}$ indicates whether the gap in control $j$ is exploited in $S_i$. \textcolor{black}{Weights $w_e=0.35$, $w_r=0.35$, and $w_c=0.30$
represent the economic, regulatory, and interaction
contributions, respectively, assigning equal importance
to the individual dimensions while limiting their
interaction term.}

\subsection{{GRID-MRI} Formulation}

The \textcolor{black}{GRID-MRI} integrates: \emph{i)} endogenous severity, \emph{ii)} non-additive coordination effects, and \emph{iii)} scenario-specific exogenous conditions. The endogenous severity gives equal weight to operational response and to combined physical and cyber exposure, defined as: \looseness=-1

\vspace{-8pt}
\begin{equation}
\bar D_i=\tfrac14\left(D_{\mathrm{phy},i}+D_{\mathrm{cyb},i}\right)
+\tfrac12D_{\mathrm{op},i}
\label{eq:severity}
\end{equation}
To capture non-additive effects, the coordinated response is compared against a linear superposition reference. Each compromised asset $b\in\mathcal B_i$ is simulated in isolation, and its rotor-speed ($\omega$) deviation from the baseline is delayed by the attack time of asset $b$ in $S_i$ before superposition:

\begin{equation}
\omega_g^{\mathrm{add}}(t)=\omega_g^0(t)
+\sum_{b\in\mathcal B_i}\!\left[\omega_g^{(b)}(t)-\omega_g^0(t)\right]
\label{eq:superposition}
\end{equation}
where $\omega_g^0(t)$ denotes the baseline $\omega$ and $\omega_g^{(b)}(t)$ the single-asset response aligned with the attack time of asset $b$ in scenario $S_i$. The resilience losses of the coordinated and additive responses, $R_{\mathrm{loss},i}$ and $R_{\mathrm{loss},i}^{\mathrm{add}}$, define the interaction factor as:

\begin{equation}
\eta_i=\frac{R_{\mathrm{loss},i}-R_{\mathrm{loss},i}^{\mathrm{add}}}
{\max\!\left(R_{\mathrm{loss},i},R_{\mathrm{loss},i}^{\mathrm{add}}\right)}
\label{eq:eta}
\end{equation}
where $\eta_i=0$ if both losses vanish or if the scenario involves a single asset, while positive values indicate that coordination amplifies the resilience loss relative to the additive reference.

Finally, the severity $\bar D_i$ is adjusted by the interaction factor $\eta_i$ and scaled by the scenario-specific climatic and economic-regulatory dimensions:

\vspace{-10pt}
\begin{equation}
\mathrm{GRID\text{-}MRI}_i=\bar D_i(1+\eta_i)(1+D_{\mathrm{clim},i})(1+D_{\mathrm{econ\text{-}reg},i})
\label{eq:mdri}
\end{equation}
where the exogenous indices are specified \textit{separately} from the dynamic simulations and act as multiplicative scaling factors.

\section{Case Study and Results}
\label{s:system_models}

\newlength{\figH}
\settoheight{\figH}{\includegraphics[trim=7.7bp 0 0.0bp 0,clip,width=0.2375\textwidth]{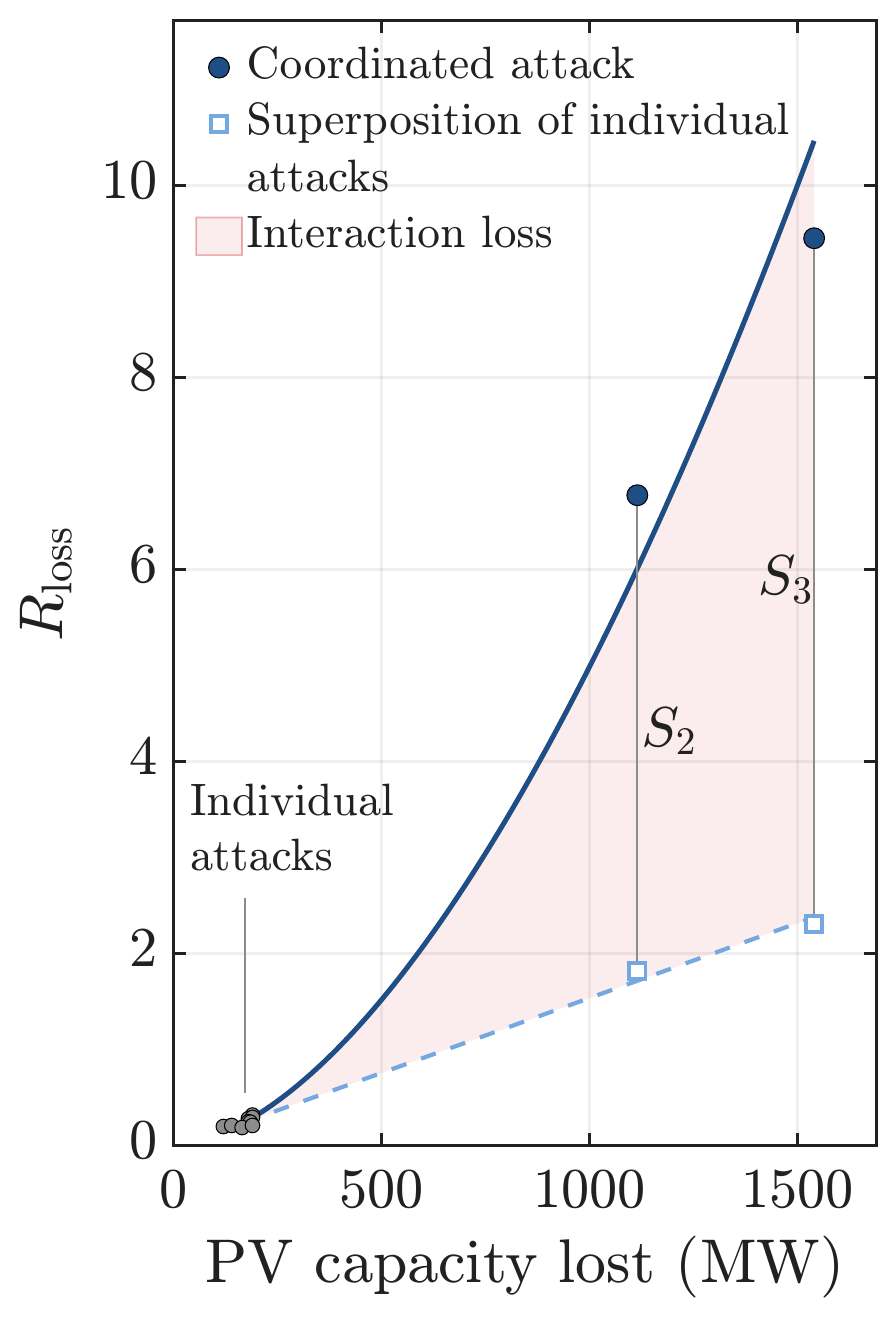}}% altura de (c)

\begin{figure*}[!t]
  \centering
  \scalebox{0.98}{%
    \begin{tabular}{@{}c@{\hspace{0.012\textwidth}}c@{\hspace{0.012\textwidth}}c@{\hspace{0.012\textwidth}}c@{}}
      \includegraphics[trim=6.7bp 0 0.0bp 0,clip,width=0.2548\textwidth]{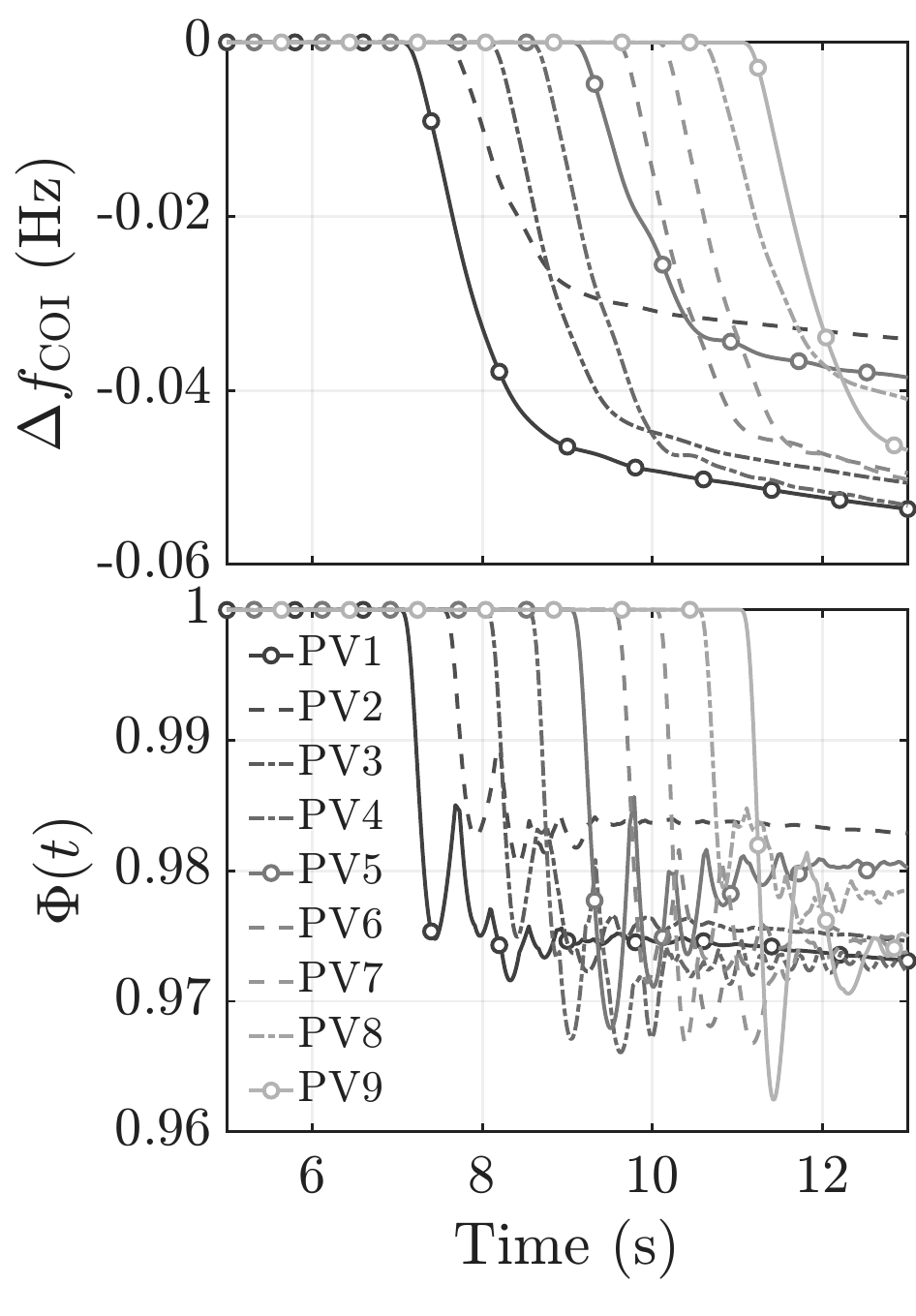} &
      \includegraphics[trim=4.3bp 0 0.0bp 0,clip,width=0.2244\textwidth]{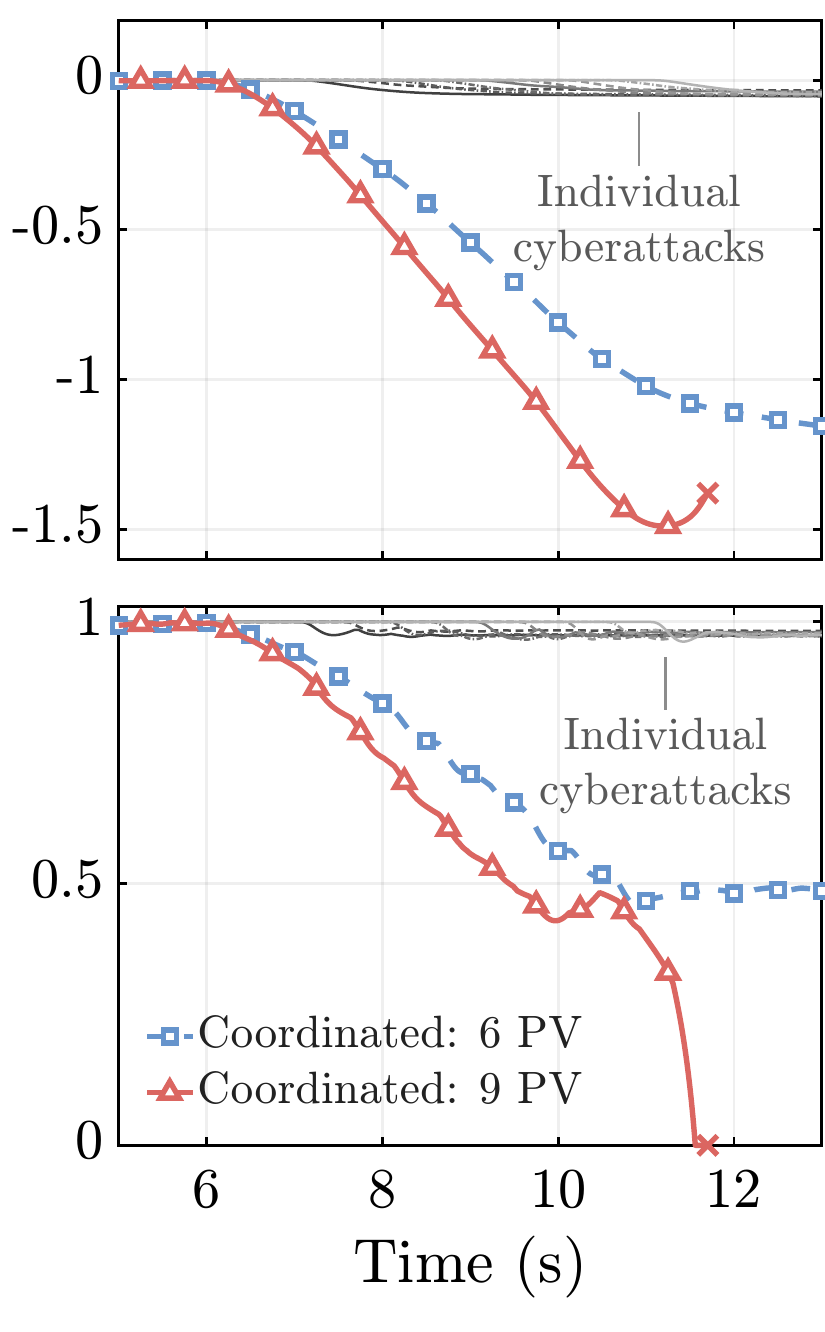} &
      \includegraphics[trim=7.7bp 0 0.0bp 0,clip,width=0.2375\textwidth]{Figures/Fig_2_2_3.pdf} &
      \raisebox{0mm}{%
        \includegraphics[
          trim=0bp 0bp 0bp 0bp,
          clip,
          height=\figH
        ]{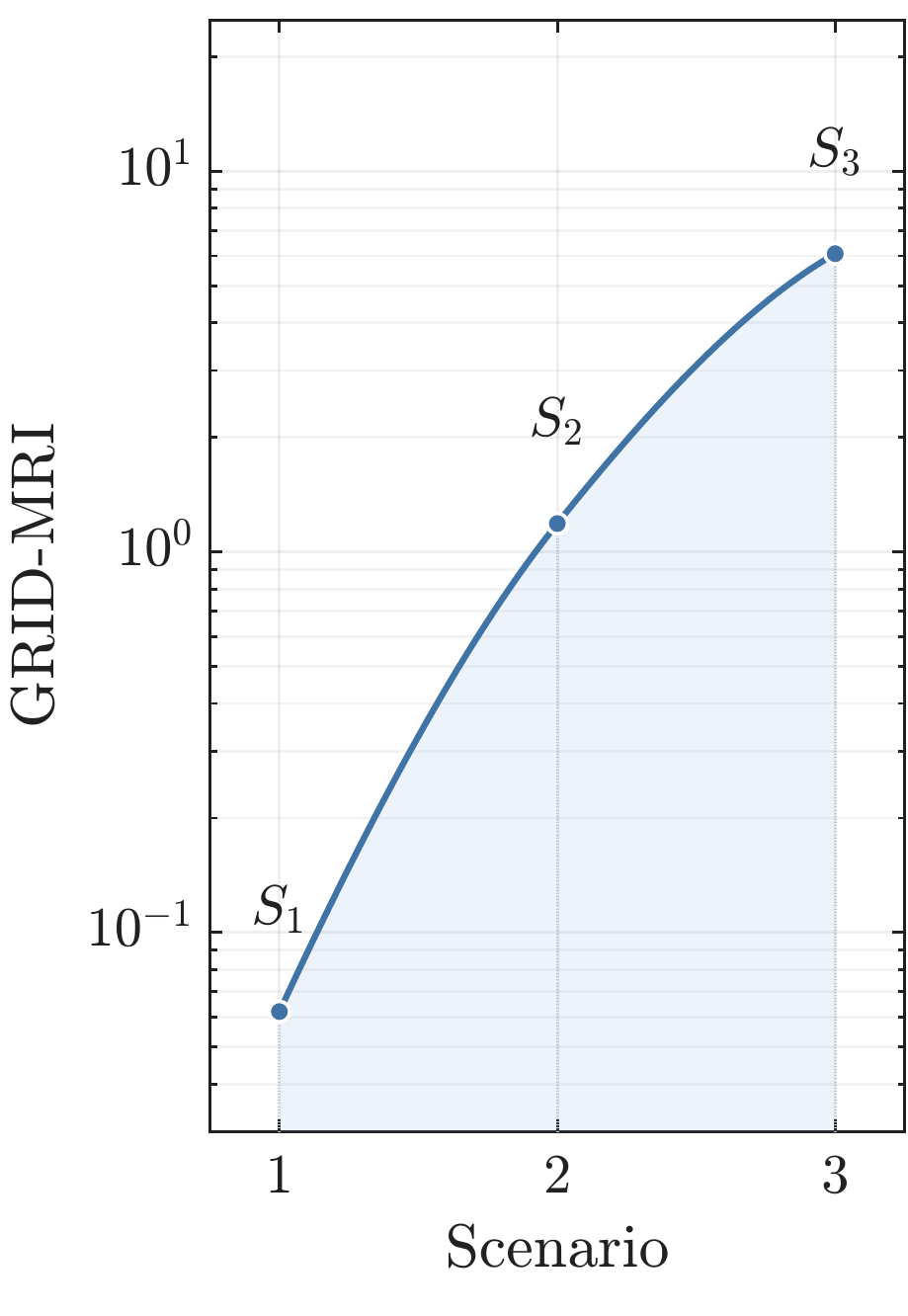}%
      } \\
      [-3mm]
      \hspace*{8mm}(a) &
      \hspace*{4mm}(b) &
      \hspace*{6mm}(c) &
      \hspace*{10mm}(d)
    \end{tabular}%
  }
  \vspace{-0.5mm}
  \caption{System response, resilience loss, and {GRID-MRI} under individual and coordinated cyberattacks.}
  \label{fig:individual_vs_coordinated}
  \vspace{-4mm}
\end{figure*}

\subsection{System and Attack Models}
The proposed framework is evaluated on the IEEE 39-bus system, implemented in MATLAB/Simulink, in which 1,540~MW of synchronous generation is replaced by nine grid-forming photovoltaic (PV) plants, corresponding to a penetration of 14.5\% and a pre-attack COI frequency of 60.015~Hz. Based on the December 2025 cyberattack on Polish energy infrastructure~\cite{CERTPolska2026,Dragos2026}, a state-sponsored adversary targeting both information technology (IT) and OT domains is assumed, and all scenarios share a 25\% winter-load increase. 

Three attack scenarios are considered: \emph{i)}~$S_1$ is a control-input attack that manipulates the active power reference of a single 190-MW plant, leaving the communication and control of the remaining plants intact, \emph{ii)}~$S_2$ is a multi-vector attack that disrupts communications with six plants, removing operator observability and control forcing disconnection and removing 1,115~MW, and \emph{iii)}~$S_3$ is a worst-case attack scenario where an adversary with unlimited resources disconnects all nine plants in a coordinated sequence, removing 1,540~MW, under additional climatic and economic-regulatory stress.

\subsection{Dynamic Response and Resilience Assessment}

Fig.~\ref{fig:scenarios} illustrates the non-attacked and attacked PV plants, $\omega(rad/s)$, and $\Phi_i(t)$ across the three scenarios, while Table~\ref{tab:freq} shows their frequency responses. In $S_1$, the loss of 190~MW at $t=8.5$s produces bounded oscillations and negligible frequency deviations, with $\Phi(t)$ reaching a minimum of 0.966 before settling at 0.973, yielding $R_{\mathrm{loss},1}=0.29$. In contrast, the coordinated disconnections in $S_2$ and $S_3$ produce progressively more severe frequency deviations and loss of $\omega$ coherency. Generator trajectories begin to diverge at approximately $t=15.5$s and $t=10.5$s, respectively, with maximum generator frequency spreads of 2.011 and 2.007~Hz. Although the maximum RoCoF remains below its critical threshold in both scenarios, the frequency nadir deviations exceed 1.0~Hz, and the generator frequency spreads indicate loss of synchronism. Consequently, $\Phi(t)$ reaches zero at 16.17s in $S_2$ and 11.56s in $S_3$, yielding resilience losses of 6.77 and 9.45, respectively. The shorter collapse time and greater resilience loss in $S_3$ demonstrate the accelerated degradation associated with the larger coordinated generation loss.

\begin{table}[t]
\footnotesize
\centering
\caption{Frequency response metrics.} \label{tab:freq} \vspace{-2mm}
\setlength{\tabcolsep}{3pt}
\resizebox{\columnwidth}{!}{%
\begin{tabular}{l c c c c}
\toprule
{Metric} & {Limit} & $S_1$ & $S_2$ & $S_3$ \\
\midrule
Pre-event frequency [Hz]              & --  & 60.015             & 60.015            & 60.015            \\
Nadir frequency [Hz]                  & --  & 59.957 & 58.801            & 58.528            \\
$|\Delta f_{\mathrm{nadir}}|$ [Hz]    & 1.0 & 0.060              & \textbf{1.215}             &\textbf{ 1.489}             \\
Time to frequency nadir [s]           & --  & 19.000 & 15.466            & 11.200            \\
$|\mathrm{RoCoF}|_{\max}$ [Hz/s]      & 1.0 & 0.044              & 0.271             & 0.395             \\
RoCoF time [s]                        & --  & 9.250              & 10.033            & 10.300            \\
$\Delta f^{\max}_{\mathrm{gen}}$ [Hz] & --  & 0.050              & 2.011             & 2.007             \\
Steady-state frequency [Hz]           & --  & 59.957 & \textbf{unstable} & \textbf{unstable} \\
\bottomrule
\end{tabular}%
}
\vspace{0.2em}

\label{tab:freq} \vspace{-5mm}
\end{table}

Table~\ref{tab:results} summarizes the disruption indices, resilience
losses, interaction factors, and resulting GRID-MRI values. The scenarios differ in the extent of the cyber compromise. $S_1$ is limited to an integrity attack at a single plant, $S_2$ compromises observability, controllability, integrity, and availability at six plants, and $S_3$ extends the compromise to all nine plants and additionally includes loss of recoverability. This escalation is reflected in the operational index, which rises from 0.052 in $S_1$ to 0.743 in $S_2$ and 0.942 in $S_3$. These results verify that coordinated multi-plant disruption produces a substantially greater operational impact than an isolated control-input attack.

\begin{table}[!t]
\footnotesize
\centering
\caption{Resilience metrics and GRID-MRI by scenario.}
\label{tab:results}
\vspace{-2mm}

\setlength{\tabcolsep}{4pt}

\begin{tabular*}{0.85\columnwidth}{@{\extracolsep{\fill}}lccc@{}}
\toprule
Metric & $S_1$ & $S_2$ & $S_3$ \\
\midrule
$D_{\mathrm{phy},i}$ & 0.123 & 0.724 & 1.000 \\
$D_{\mathrm{cyb},i}$ & 0.022 & 0.533 & 1.000 \\
$D_{\mathrm{op},i}$ & 0.052 & 0.743 & 0.942 \\
$D_{\mathrm{clim},i}$ & 0 & 0 & 0.783 \\
$D_{\mathrm{econ\text{-}reg},i}$ & 0 & 0 & 1.000 \\
$\bar D_i$ & 0.062 & 0.686 & 0.971 \\
$R_{\mathrm{loss},i}$ & 0.29 & 6.77 & 9.45 \\
$R_{\mathrm{loss},i}^{\mathrm{add}}$ & 0.29 & 1.82 & 2.31 \\
$\eta_i$ & 0 & 0.732 & 0.756 \\
\midrule
$\mathrm{\textbf{GRID\text{-}MRI}}$ & 0.062 & 1.188 & 6.081 \\
\bottomrule
\end{tabular*}

\vspace{-5mm}
\end{table}

The climatic and economic-regulatory dimensions are excluded from $S_1$ and $S_2$ (i.e., $D_{\mathrm{clim}}=D_{\mathrm{econ\text{-}reg}}=0$),
so that their endogenous disruption mechanisms can be evaluated independently of exogenous amplification. In $S_3$, the four equally weighted stressors (i.e., temperature, snow, wind, and ice) have normalized intensities $I_{c_k,3}$, yielding $D_{\mathrm{clim},3}=0.783$. \looseness=-1

Following the system collapse in $S_3$, the winter-adjusted demand of $\approx 7{,}600$~MW is assumed to be interrupted and linearly restored over 18~h, inspired by the restoration following the 28 April 2025 Iberian blackout~\cite{entsoe2026gridincident}. This results in an unserved energy of $\mathrm{ENS}_3=68{,}400$~MWh,
equivalent to \texteuro~547.2~million at a VoLL of \texteuro~8,000/MWh~\cite{Schroder2015,NIS2Directive2022,
IRENA2024,ENISA2023,ENTSOE2023}.
Adding \texteuro~31.5~million in equipment replacement, \texteuro~10~million
in regulatory penalties, and \texteuro~5.5~million in islanding costs yields
a total cost of $C_3^{\mathrm{tot}}=\text{\texteuro~594.2}$~million. Since this
exceeds the minimum preventive investment of \texteuro~10.8~million,
$D_{\mathrm{econ},3}^{\mathrm{und}}\approx1$~\cite{IEC62443,IRENA2024,ENISA2023,ENTSOE2023}. Assuming that all reference
controls are exploited ($D_{\mathrm{reg},3}^{\mathrm{vul}}=1$), the combined
index reaches $D_{\mathrm{econ\text{-}reg},3}=1.000$.

\textcolor{black}{Fig.~\ref{fig:individual_vs_coordinated} compares the system response under individual and coordinated cyberattacks. As shown in Fig.~\ref{fig:individual_vs_coordinated}~(a), individual attacks produce limited frequency deviations and bounded performance oscillations, preserving system stability, with a maximum resilience loss of 0.32. In contrast, Fig.~\ref{fig:individual_vs_coordinated}~(b) shows that coordinated attacks induce substantially larger frequency deviations and progressive performance degradation, ultimately leading to system collapse (red and blue lines).}

\textcolor{black}{Fig.~\ref{fig:individual_vs_coordinated}~(c) quantifies this effect by comparing the resilience losses of coordinated attacks against the additive superposition of their individual attack responses. While the additive references increase approximately linearly with the compromised PV capacity, coordinated attacks exhibit nonlinear growth, producing resilience losses 3.72 and 4.09 times greater than their respective additive estimates in $S_2$ and $S_3$. The resulting interaction factors, $\eta_2=0.732$ and $\eta_3=0.756$, quantify the additional degradation attributable to attack coordination.}

\textcolor{black}{Fig.~\ref{fig:individual_vs_coordinated}~(d) illustrates the increase in GRID-MRI from 0.062 in $S_1$ to 1.188 in $S_2$ and reaches 6.081 in $S_3$. The logarithmic decomposition reveals that the initial increase is predominantly driven by endogenous disruption severity (81\%), whereas the subsequent increase is primarily associated with climatic and economic-regulatory amplification (78\%). Thus, coordinated disturbances govern the transition toward system collapse, while exogenous stressors further increase the GRID-MRI under the worst-case conditions.}

\section{Conclusions}
\label{s:Conclusions}

This work proposes GRID-MRI, a metric that integrates disruption severity, coordination effects (quantified through counterfactual simulations), and exogenous stressors to assess power system resilience under disturbances. The IEEE 39-bus case study demonstrates that coordinated cyberattacks induce system collapse despite stable individual and additive responses, producing resilience losses 3.72 and 4.09 times greater than their respective additive estimates. These findings reveal the limitations of additive assessments in capturing nonlinear degradation induced due to attack coordination.

Climatic and economic-regulatory conditions further amplify the assessed resilience impact under the worst-case scenario, accounting for a 78\% increase from $S_2$ to $S_3$. By distinguishing the contributions of endogenous severity, coordination effects, and exogenous amplification, the proposed  GRID-MRI metric provides a comprehensive characterization of resilience degradation and supports the identification of dominant vulnerabilities under coordinated disturbances.

\section*{Acknowledgment} %\vspace{-1mm}
This work is funded by the National Science Foundation (NSF) Award Number \#2501975.
\vspace{-1mm}

\small
\bibliographystyle{elsarticle-num} 
\bibliography{biblio.bib}

\end{document}